\documentclass[11pt]{article} 

\usepackage{arxiv}
\usepackage[utf8]{inputenc}
\usepackage[T1]{fontenc}

\usepackage[hidelinks]{hyperref} 
\usepackage{graphicx}
\usepackage{url}
\usepackage{natbib}
\usepackage{color} 
\usepackage{multirow}
\usepackage{amssymb}
\usepackage{amsmath}
\usepackage{bbm, dsfont}
\usepackage{amssymb}
\usepackage{theorem}
\usepackage[noend]{algpseudocode}
\usepackage[linesnumbered,ruled,vlined]{algorithm2e}
\usepackage{algorithmicx}

\theoremstyle{plain}
\theorembodyfont{\rmfamily}

\newtheorem{dfn}{Definition}

\newtheorem{remark}{Remark}
\newtheorem{prop}{Proposition}

\usepackage{tikz}

\usepackage[english]{babel}
\addto\captionsenglish{}

\begin{document}

\title{Point Pattern Processes and Models} 

\author{
	Nik Lomax \thanks{Corresponding author: \textit{N.M.Lomax@leeds.ac.uk}}
	 \\ Leeds Institute for Data Analytics\\ University of Leeds\\ Leeds, LS2 9JT, UK
	\And
	Nick Malleson\\ School of Geography\\  University of Leeds\\ Leeds, LS2 9JT, UK
	\And
	Le-Minh Kieu\\ Leeds Institute for Data Analytics\\ University of Leeds\\ Leeds, LS2 9JT, UK 
}

\maketitle

\begin{abstract}
In recent years there has been a substantial increase in the availability of datasets which contain information about the location and timing of an event or group of events and the application of methods to analyse spatio-temporal datasets spans many disciplines. This chapter defines and provides an overview of tools for analysing spatial and temporal point patterns and processes, where discrete events occur at random across space or over time respectively. It also introduces the concept of spatial-temporal point patterns and methods of analysis for data where events occur in both space and time. We discuss models, methods and tools for analysing point processes. 
\end{abstract}

\tableofcontents 


\section{Introduction}

In recent years there has been a substantial increase in the availability of datasets which contain information about the location and timing of an event or group of events. Examples commonly cited include the occurrence of wild fires or earthquakes, the timing and location of crimes or the spread of disease in an epidemic. Thorough worked examples are provided in \citet{diggle_spatiotemporal_2006}, including the occurrence of gastro-intestinal illness in humans and the spread of bovine tuberculosis. Alongside this rise in data availability, \citet{gonzalez_spatiotemporal_2016} points to an acceleration in methodological developments over the last decade which can be used to analyse patterns of points in space and time. Entire volumes have been dedicated to the consideration of point processes, see for example \textit{An Introduction to the Theory of Point Processes} by \citet{daley_introduction_2003} which provides a history of point process analysis and detailed consideration of underlying processes. The application of methods to analyse spatio-temporal datasets spans many disciplines. 

Spatial point processes and temporal point processes are stochastic processes, where discrete events occur at random across space or over time respectively. The number of points (events) is also random. By extension, spatial-temporal data provide information about the dispersion of events across both space {\it and} time. In this case the basic format of these data is $(s_i, t_i): i=1, \ldots, n$ where each $s_i$ denotes the location and $t_i$ the corresponding time of occurrence of an event i of interest. Assuming that the data form a complete record of all events that occur within a pre-specified region R (usually 2-dimensional) and a pre-specified time interval (0, T), a data set of this form may be called a spatial-temporal point pattern (also called space-time or spatio-temporal point pattern). The underlying stochastic model for the data is a spatial-temporal point process. An excellent text on spatio-temporal point processes is provided by \citet{diggle_spatiotemporal_2006}, who defines the stochastic process as a countable set of points (events) in some pre-defined space, and the pattern as a finite set of points treated as a partial realisation of a stochastic process.

While there is extensive literature on the analysis of point process data in space or in time, \citet{gabriel_stpp_2013} argue that the literature on generic methods for analysis of spatial-temporal point processes is less well established. However, many of the tools used to analyse spatial (or temporal) point pattern data can be extended to the spatial-temporal setting and the spatial-temporal setting opens up analysis/modelling strategies that more explicitly account for the directional character of time.  Spatio-temporal methods for analysis are required where separate analyses of the spatial component and the temporal component of the data are not appropriate. \citet{diggle_spatiotemporal_2006} points out that when extending to spatio-temporal patterns and processes it is useful to allow either the spatial or temporal dimensions to be discrete (these situations are termed respectively spatially discrete or temporally discrete). Situations where neither the spatial or the temporal are discrete are termed continuous spatio-temporal point processes.

In this chapter, the next section is dedicated to defining the basic concepts and statistical methods for spatial point patterns, with a note about extension to temporal point patterns. We then provide an overview of models which can be applied to point processes, including Homogeneous Poisson, Non-homogeneous Poisson and Hawkes Processes. Examples of temporal point processes are used to exemplify these models. The subsequent section provides discussion of some methods and tools for analysing and visualising point patterns, with a focus on spatial examples and an extension to spatial temporal methods.  Finally we offer some conclusions.
\section{Basic Concepts for Spatial Point Patterns}

\begin{figure}[] \begin{center}
		\resizebox{1.0\linewidth}{!}{
			\includegraphics[]{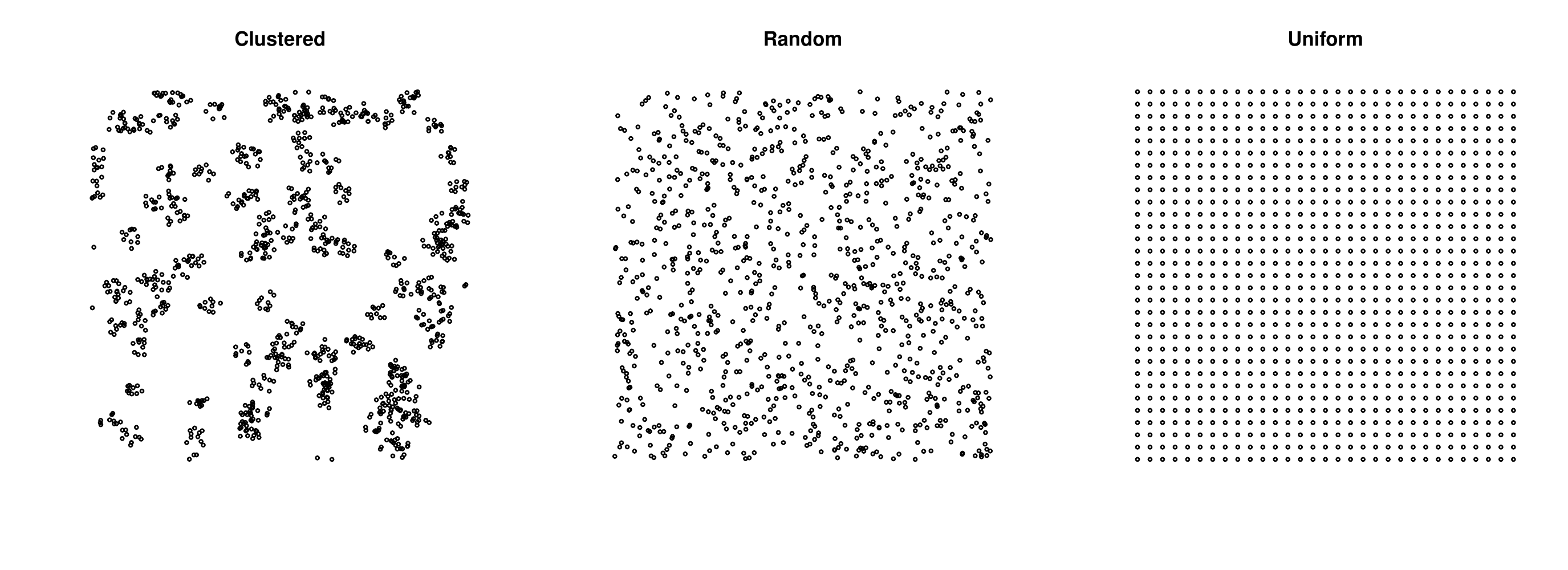}
		}
		\caption{Point patterns showing (a) random point samples, (b) clustered point samples and (c) regular point samples.}
		\label{fig:point_pattern_randomness} \end{center} \end{figure}

For a thorough and detailed text on spatial point patterns and processes the reader is directed to \cite{diggle_statistical_2013}, and a second reference to note is \cite{bivand_applied_2008} who use a range of R packages to implement examples discussed in \cite{diggle_statistical_2013}. Much of the discussion in this section refers to these key references. A set of points distributed across a region can be described as a spatial point pattern. Those points can be defined as cases, and each case can have information attached to it (e.g. a species of tree).  The distribution of these event data across a region can be defined broadly by three types of pattern, demonstrated in Figure~\ref{fig:point_pattern_randomness}. These are:
\begin{itemize}
	\item Clustered, or aggregated -- where there is some underlying process to the clustering
	\item Completely random -- where there is no pattern to the distribution
	\item Regular –- where events are fairly evenly distributed
\end{itemize}

The clustered pattern demonstrates that certain areas have a higher than average number of events. Clustering usually occurs where there is attraction between points. The regular pattern demonstrates an equal distribution of events across space, which occurs where there is inhibition (i.e. competition) between points.

A completely random spatial pattern requires the average number of events to be homogeneous, the events to be independent and for the events to follow the Poisson distribution. A random pattern demonstrates no obvious structure and points are independent from one another. The starting point for much analysis of spatial point patterns is to test for Complete Spatial Randomness (CSR).  \cite{diggle_statistical_2013} argues that a pattern for which the hypothesis of CSR is not rejected merits little further formal statistical analysis and as such CSR deserves further consideration here.

\subsection{Complete Spatial Randomness (CSR)}

Complete spatial randomness (CSR) describes a process whereby events occur within a given study area (R) completely at random. It is synonymous  with a homogeneous Poisson process, which we discuss later in this chapter. CSR requires the average number of events to be homogeneous, the events to be independent and for the events to follow the Poisson distribution. A random pattern demonstrates no obvious structure and points are independent from one another. \cite{diggle_statistical_2013} notes that testing for complete spatial randomness should be the starting point of any framework of analysis of spatial point patterns as this is an important condition. CSR is a frame of reference against which other patterns can be tested. As defined by \cite{diggle_statistical_2013}, tests for CSR include:

\begin{itemize}
\item Distance to the Nearest Event. The G function measures the distribution of the distances from an arbitrary event to its nearest event.
\item Distance from a Point to the Nearest Event. 
The F function measures the distribution of all distances from an arbitrary point of the plane to its nearest event. 
\item Quadrat counts, where the spatial plane is dissected in to sub-regions representing quadrats.  Points are counted in each quadrat and the intensity is compared.
\item Scales of pattern, where an index of dispersion (calculated as the the sample ratio to mean) is calculated for various grids constructed from adjacent quadrats. These grids are defined as \textit{blocks} and the index of dispersion is plotted against the size of the block. Peaks or troughs in the relationship on the graph are interpreted as evidence of scales of pattern where peaks represent aggregated and troughs represent regular patterns.
\end{itemize}

We expand on distance measures later in this chapter. Testing for CSR is not in itself of particular interest, rather it can assist in the formulations of alternatives to CSR.

\subsection{Spatial Point Processes}
As defined by \cite{diggle_spatiotemporal_2006} Spatial Point Processes are stochastic mechanisms which generate a series of events across a region. The locations of the events generated by a point process are a point pattern. Spatial Point Processes are generally static, whereby the statistical parameters of the underlying process do not vary over space (invariant under translation), and isotropic, whereby that they exhibit the same value when measured from different directions (invariant under rotation). Later in this chapter  we discuss methods of analysis for point processes, including Homogeneous, non-Homogeneous and Hawkes processes. One can define first order properties and second order properties of the spatial point process. 

\subsection{First and Second order properties}

First-order properties relate to intensity and spatial density over a study region, defined as the mean number of events per unit area. Neither the intensity nor the density provide any information about the interaction between points. Analysis of first-order properties commonly involves the use of intensity estimators, for example kernel density estimation (expanded upon later in this chapter). Kernal estimates can be distorted by \textit{edge effects}, which occur where points are close to the boundary of the region being observed. The problem, summarised by \cite{diggle_statistical_2013} is that events outside the region of study may be interacting with events inside the region, however it is difficult to take account of this. 

Second-order properties relate to the strength and type of interactions between events of the point process, and involve a relationship between numbers of events in pairs of areas. As defined by \cite[p.~171]{bivand_applied_2008} ``the second-order intensity of two points \emph{x} and \emph{y} reflects the probability of any pair of events occurring in the vicinities of \emph{x} and \emph{y}, respectively." Second order properties can be estimated using the K function, which measures the number of events found up to a given distance from any particular event. See discussion of Ripley's-K later in this chapter for further information. 

\subsection{Extension to Temporal data}

So far in this section we have discussed spatial point processes, but methods can easily be expanded to take in to account temporal data. A Point Process often refers to a mathematical construct to model the times at which event happens, which we shall denote by $T_1,T_2, \ldots$. For example, if we use a Point Process to model customers arriving to a supermarket, $T_1$ represents the time when customer 1 arrives at the shop, $T_2$, represents the following customer arrival and so on. 
$T_k$ can usually be interpreted as the time of occurrence of the $k$-th event, in this case - the $k$-th arrival. In this chapter henceforth, we refer to $T_i$ as event times. 

The number of events in a Point Process is often referred as a counting process $N_t$ is a random function defined on time $t\geq 0$, and take integer values $1,2,\ldots$. 
Its value is the number of events occurring within time 0 to $t$. Therefore it is uniquely determined by a sequence of non-negative random variables $T_i$, satisfying $T_i<T_{i+1}$ if $T_i\leq\infty$. $N_t$ {\em counts} the number of events up to time $t$, i.e. ${N}_t:=\sum_{i\geq 1}\mathbbm{1}_{\{t\geq T_i\}}$, where  $\mathbbm{1}_{\{\cdot\}}$ is the indicator function that takes value 1 when the condition is true and 0 otherwise. We can see that $N_0=0.$ $N_t$ is piecewise constant and has jump size of 1 at the event times $T_i$. 

\begin{figure}[] \begin{center}
		\resizebox{0.5\linewidth}{!}{
			\includegraphics[]{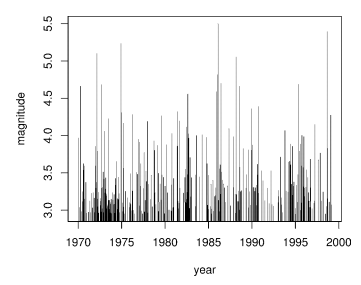}
		}
		\caption{The times and magnitude of earthquakes in Bear Valley, California. Source: \cite{schoenberg_multidimensional_2003}}
		\label{fig:earthquakes} \end{center} \end{figure}

The distribution of event data over time is considered a temporal point process if events are random. For example, Figure~\ref{fig:earthquakes}, taken from \cite{schoenberg_multidimensional_2003} shows the timing and magnitude of earthquakes in Bear Valley, California between 1970 and 2000. No obvious trend in the magnitude distribution over time is easily discernible. Note that there is a well established literature for the modelling of earthquakes using spatial-temporal methods.

\section{Models}
\label{s:models}
In this section we provide an overview of a number of models which can be applied to point processes: Homogeneous Poisson processes; Non-homogeneous Poisson processes and Hawkes processes. We then discuss the simulation of these processes to generate pseudo-random data points from them. The examples used in this section relate to temporal point processes, however these models could equally be applied to spatial point processes.

\subsection{Homogeneous Poisson Process}

The simplest class of point process is the \textit{Homogeneous Poisson process} (HPP). It is a point process of the occurrence of random events in continuous time $T_i$, $i = 1,2,3\dots$, which satisfy the following assumptions:

\begin{itemize}
	\item Events occur singly: during a short time interval, the probability of
	observing more than one event is negligible
	\item Inter-arrival
	times between events are assumed to be independent of one
	another
	\item The occurrence rate of event remains \textit{constant} within the modelling period
\end{itemize}

An example of a Poisson Process can be a model of light bulb failures in a big building. They rarely happen simultaneously and they may occur at a constant rate. The Poisson Process model may give an indicator of the number of failures within a study period to prepare a stock of light bulbs. Because the occurrence rate is constant, this process is often referred as Homogeneous Poisson Process (HPP). Formally, it can be defined as follows:

\begin{dfn} \label{def:poisson-process}
	\textbf{Poisson process:}
	Let $(Q_k)_{k\geq 1}$ be a sequence of independent and identically distributed Exponential random variables with parameter $\lambda$ and event times $T_n=\sum_{k=1}^{n}\,Q_i$. 
	The process $(N_t\,,\,t\geq 0)$ defined by ${N}_t:=\sum_{k\,\geq 1\,}\mathbbm{1}_{\{t\geq T_k\}}$ is called a \textit{Poisson process} with intensity $\lambda$.
\end{dfn}

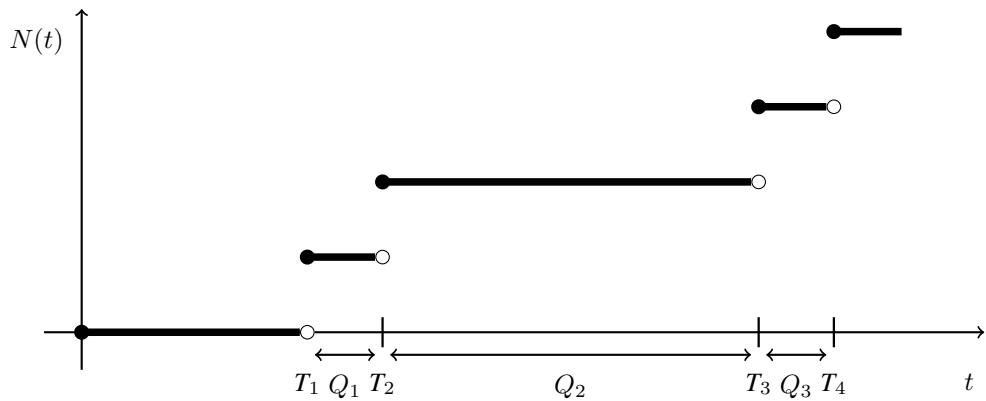
\begin{figure}[h]
	\centering
	\begin{tikzpicture}
	\tikzstyle{vertex}=[circle, draw=black,fill = white, line width=1.8mm, minimum size=1pt, inner sep=0pt]
	\tikzstyle{vertex1}=[circle, draw=black, fill=black, line width=1mm, minimum size=1pt, inner sep=0pt]
	\tikzstyle{edge} = [draw, line width=1mm, -]
	\tikzstyle{edge1} = [draw, line width=0.3mm, -]
	
	\node[vertex1,label=below:{},inner sep=1pt,minimum size=1pt] () at (0,0) {};
	\node[label=below:{$\textit{T}_1$},inner sep=0pt,minimum size=1pt] (t1) at (3,-0.4) {};
	\node[circle,draw=black,label=below:{},inner sep=0pt,minimum size=1.8mm] (t1) at (3,0) {};
	\node[label=below:{$\textit{T}_2$},inner sep=0pt,minimum size=1pt] (t1) at (4,-0.4) {};
	\node[circle,draw=black,label=below:{},inner sep=0pt,minimum size=1.8mm] (t1) at (4,1) {};
	\node[label=below:{$\textit{T}_3$},inner sep=0pt,minimum size=1pt] (t1) at (9,-0.4) {};
	\node[circle,draw=black,label=below:{},inner sep=0pt,minimum size=1.8mm] (t1) at (9,2) {};
	\node[label=below:{$\textit{T}_4$},inner sep=0pt,minimum size=1pt] (t1) at (10,-0.4) {};
	\node[circle,draw = black,label=below:{},inner sep=0pt,minimum size=1.8mm] (t1) at (10,3) {};
	\node[vertex1,label=below:{},inner sep=1pt,minimum size=1pt] (t1) at (3,1) {};
	\node[vertex1,label=below:{},inner sep=1pt,minimum size=1pt] (t1) at (4,2) {};
	\node[vertex1,label=below:{},inner sep=1pt,minimum size=1pt] (t1) at (9,3) {};
	\node[vertex1,label=below:{},inner sep=1pt,minimum size=1pt] (t1) at (10,4) {};
	\node[label=below:{$t$},inner sep=0pt,minimum size=1pt] (t1) at (11.8,-0.4) {};
	\node[label=below:{$N(t)$},inner sep=0pt,minimum size=1pt] (t1) at (-0.6,4.2) {};
	\node[label=below:{$Q_1$},inner sep=1pt,minimum size=1pt] () at (3.5,-0.4) {};
	\node[label=below:{$Q_2$},inner sep=1pt,minimum size=1pt] () at (6.5,-0.4) {};
	\node[label=below:{$Q_3$},inner sep=1pt,minimum size=1pt] () at (9.5,-0.4) {};
	
	\draw[thick,-] (-0.5,0) -- (2.9,0);
	\draw[thick,->] (3.1,0) -- (12,0);
	\draw[thick,->] (0,-0.5) -- (0,4.3);
	\draw[thick,<->] (3.1,-0.3) -- (3.9,-0.3);
	\draw[thick,<->] (4.1,-0.3) -- (8.9,-0.3);
	\draw[thick,<->] (9.1,-0.3) -- (9.9,-0.3);
	\path[edge] (0,0) -- (2.9,0);
	\path[edge] (3,1) -- (3.9,1);
	\path[edge] (4,2) -- (8.9,2);
	\path[edge] (9,3) -- (9.9,3);
	\path[edge] (10,4) -- (10.9,4);
	\path[edge1] (4,0.2) -- (4,-0.2);
	\path[edge1] (9,0.2) -- (9,-0.2);
	\path[edge1] (10,0.2) -- (10,-0.2);
	
	\end{tikzpicture}
	\caption{
		A sample path for Poisson process with four observed events.
		The y-axis depicts $N(t)$, while the x-axis represents time.
		$T_i$ represent event times, while $Q_j$ represent inter-arrival times.}
	\label{fig:poisson-process}
\end{figure}	

The sequence of $Q_k$ are known as the \emph{inter-arrival times} (as in Figure \ref{fig:poisson-process}), and it can be interpreted as follows in terms our example on light bulb failures: the first bulb fails at time $Q_1$, the second fails at $Q_2$ after the first, and so on. One can show that this construct means that light bulb fails at an average rate of $\lambda$ per unit time, since the expected time between event times is $\frac{1}{\lambda}$. 
The number of events which occur in the half-open interval $(a_i,b_i]$ with $a_i < b_i \leq a_{i+1}$ is estimated as: 

\begin{equation}
Pr\{ N(a_i,b_i] = n_i, i=1,...,k \} = \prod_{k}^{i=1} \frac{[\lambda (b_i-a_i)] ^ {n_i}}{n_i!} e^{-\lambda(b_i-a_i)}
\label{eq:Poisson_count}  
\end{equation}

Equation \ref{eq:Poisson_count} shows that the count of events is, in fact, Poisson distributed. This leads to the fact that the mean $M(a,b]$ and variance $V(a,b]$ of the number of events within the interval $(a_i,b_i]$ is equal. 

\begin{equation}
M(a,b] = \lambda(b-a)= V(a,b]
\label{eq:mean_var_Poisson}
\end{equation}

The parameter $\lambda$ can be interpreted as the mean rate of the HPP. 
Assume that $r$ time units have elapsed and during this period no events have arrived, i.e. there are no events during the time interval $[0,r]$. 
The probability that we will have to wait a further $t$ time units given by
\begin{align} \label{eq:memorylessness}
\qquad\qquad p (Q>t+r\,|\,Q>r)&=\frac{p(Q>t+r\,,\,Q>r)}{p(Q>r)} \nonumber \\
\qquad\qquad&=\frac{p(Q>t+r)}{p(Q>r)}=\frac{\exp({-\lambda(t+r)})}{\exp({-\lambda r})}=\exp({-\lambda t})=p(Q>t).
\end{align}

This can also be interpreted as the probability of having an amount of time $\tau$ between events. Equation \ref{eq:memorylessness} therefore shows that the interval has an exponential distribution with mean 1/$\lambda$. 
Suppose we were waiting for an occurrence of an event, say another bulb needs replacement, the inter-arrival times of which follow an Exponential distribution with parameter $\lambda$. 
Eq.~\eqref{eq:memorylessness} is said to have no memory and it is a special property of the HPP. 

\begin{remark} \label{remark:Memorylessness}
	\textbf{Memorylessness in Poisson Process} the likelihood and chance to wait an additional $t$ time units after already having waited $m$ time units is the same as the probability of having to wait $t$ time units when starting at time $0$. 
\end{remark}

Putting it differently, if one interprets $Q$ as the time of arrival of an event where $Q$ follows an Exponential distribution, the distribution of $Q-m$ given $Q>m$ is the same as the distribution of $Q$ itself.

\subsection{Non-homogeneous Poisson Process}

The HPP is simply characterised by a \textit{constant} arrival rate $\lambda$. It is equivalent to an assumption, for example, light bulbs fail at the constant rate, or customers arrive at a constant rate at shops at any time of day or season. This assumption may not hold in practice, because light bulbs may fail more during night time due to heavier usage, or customers may arrive more often to shops during peak periods. It is therefore useful to extend the Poisson process to a more general Point Process in which the arrival rate varies as a function of time. This type of process is often referred as non-homogeneous Poisson process (NHPP). 

\begin{dfn}
	A Point Process $N_t$ is considered as a non-homogeneous Poisson process with intensity
	function $\lambda (t)$ if it satisfies the following
	conditions: 
	
	\begin{enumerate}
		\item $P(N(0)=0)=1$
		
		\item For any interval, if the point process $N$ is a NHPP with intensity function $\lambda(t)$, then $N(t)$ follows a Poisson distribution with parameter $\mu=\int _{t_1}^{t_2}\lambda (t)dt$. 
		
		\item The number of points in the interval $[s,t]$ follows a Poisson distribution with parameter $\int _{t_1}^{t_2}\lambda (t)dt$, i.e.
		$ p(N_{t_2}-N_{t_1}=n)=\frac{1}{n!}\cdot\exp\left(-\int _{t_1}^{t_2}\lambda (t)dt\right)\left(\int _{t_1}^{t_2}\lambda (t)dt   \right)^n$. 
		\item  For any non-overlapping intervals, $(t_1,t_2]$
		and $(t_3,t_4]$, $ N_{(t_1,t_2]} $
		and $ N_{(t_3,t_4]}$ are   independent.   
	\end{enumerate} 
	
\end{dfn}

This process is called an non-homogeneous Poisson
process (NHPP) because it holds Poisson increments but the mean does change in each increment, depends on the rate function $\lambda(t)$ over the study interval. While NHPP is no longer static (as in HPP), the increments remain independent and memoryless (as in Remark \ref{remark:Memorylessness} ). A NHPP with rate function $\lambda(t)=\lambda$ for each $t \leq 0$ will reduce to a HPP. Otherwise, one dimensional distribution of NHPP is given by the equation:

\begin{equation}
P(N(t)=k)= \frac{(\int_{0}^{t}\lambda(u)du)^k}{k!} \ e^{-\int_{0}^{t}\lambda(u)du}  \quad k=0,1,2,...
\end{equation}

The expectation and variance of NHPP are given as

\begin{align}
\Lambda(t) = E[N(t)] = \int_{0}^{t}\lambda(u)du, \quad t \geq 0 \\
V(t) = V[N(t)] = \int_{0}^{t}\lambda(u)du, \quad t \geq 0 
\end{align}

Thus the expected value of increment is 

\begin{equation}
E[N(t+h)-N(t)]= \int_{0}^{t}\lambda(u)du
\end{equation}

\subsection{Hawkes Process}

In a one dimensional lattice, a Hawkes process, similar to other point processes, is a specific distribution of points on a line. Similarly, a spatial Hawkes process, on a two-dimensional spatial area, specifies points as events. The major difference between a Hawkes process compared to a Poisson process is the fact that events are correlated. The memorylessness property (in Remark \ref{remark:Memorylessness}) is no longer valid as a result. After an event, the intensity shoots up (\textit{selfexcite}) and then forgets its past (\textit{decays}). 

Formally, a Hawkes process, also known as a self-exciting process, is a point process where the arrival of 
an event causes the intensity function to increase \citet{hawkes_spectra_1971}. 

\begin{dfn} \label{dfn:Hawkes}
	A point process $N_t$ is considered as a Hawkes process if the conditional intensity function $\lambda(t|\mathcal{H}_t)$ takes the form: 
	
	\begin{equation}
	\lambda(t|\mathcal{H}_t) = \lambda_0(t) + \sum_{i:t>T_i}^{} \phi (t-T_i)
	\end{equation}
	
\end{dfn}

Where $T_i<t$ are the times for all events occurred before time $t$. These events contribute to the conditional intensity at time $t$. $\mathcal{H}_t$ is the associated history at time $t$. 

In Definition \ref{dfn:Hawkes}, $\lambda_0(t): \mathbb{R} \rightarrow \mathbb{R}_+$ is the baseline intensity. It describes the arrival of events triggered by external sources. The events are independent on the previous events. $\phi:\mathbb{R} \rightarrow \mathbb{R}_+$ is the memory kernel, or also referred as the \textit{decay function} of the Hawkes process. The kernel $\phi (t-T_i)$ decides how an event at time $T_i$ triggers a change on the intensity function at time $t$. Function $\phi(\dot)$ is usually monotonically decreasing, so that more recent events have higher impact on the current intensity. A popular decay function is the exponential function, as proposed by \citet{hawkes_spectra_1971}: 

\begin{equation}
\phi(x) = \alpha e^{-\delta x}
\end{equation}

where $\alpha \geq 0, \delta > 0$ and $\alpha < \delta$. Another popular decay function is the power-law function \citep{ozaki_maximum_1979}: 

\begin{equation}
\phi(x) = \frac{\alpha}{(x+\delta)^{\eta+1}}
\end{equation}

where $\alpha \geq 0, \delta, \eta > 0$ and $\alpha < \delta$.

The Hawkes process can be considered as an extension of the NHPP, in which the intensity function is stochastic and explicitly depends on historical events through the memory kernel $\phi(\dot)$. Therefore, events in a Hawkes process can also be classified into two groups: \textit{immigrants} and \textit{offspring}. The immigrant events are independently dictated by $\lambda_0(t)$, which itself can be modelled by a NHPP. The offspring events are only triggered by occurred events, and thus each have a parent event. 

An example of such textit{immigrants} and \textit{offspring} events in a Hawkes process is a model that can simulate social media contributions. There are social media contributions, such as Tweets, come independently from individual life events (\textit{immigrants}). Contributions by famous people, or a big news from media, may trigger a chain of contributions (\textit{offspring}) among the people who follow the original contribution. This complex phenomena can be modelled by a Hawkes process. 

The separation of events into \textit{immigrants} and \textit{offspring} is called the \textit{branching structure} of the Hawkes process. An important definition related to the branching structure is the \textit{branching factor}, which can be defined as follows: 

\begin{dfn}
	\textbf{Branching factor:} A branching factor $n^*$ is the expected number of direct offspring triggered by a single event. It can be estimated as follows: 
	
	\begin{equation}
	n^* = \int_{0}^{\inf} \phi (t) dt
	\end{equation}
	
\end{dfn}

The event that triggers an offspring is called the  \textit{immediate parent} of the offspring. All the events which are directly or indirectly triggered to an immigrant form a cluster of offspring events. The branching factor indicates whether the number of offspring is bounded or unbounded.  
If $n^* < 1$, the Hawkes process is called a \textit{subcritical regime}, because the total events triggered from a parent event is bounded, and we can estimate the size of each cluster more accurately. If $n^* > 1$, the process is called a \textit{supercritial regime}, because the number of events triggered is unbounded. Let $E_i$ be the expected number of events triggered from an immediate parent, and $E_0 =1$, which means that each cluster has been originally triggered from a single immigrant. We then can estimate the number of total event as follows: 

\begin{equation}
N_\infty = \sum_{i=0}^{\infty} A_i = \sum_{i=0}^{\infty} A_{i-1} n^* = \sum_{i=0}^{\infty} A_{i-2} (n^*)^2 = \sum_{i=0}^{\infty} (n^*)^i \quad ,i \geq 1
\end{equation}

As discussed, estimating $N_\infty$ becomes easier if $n^*<1$: 

\begin{equation}
N_\infty = \sum_{i=0}^{\infty} A_i = \frac{1}{1-n^*} \quad ,n^* < 1
\end{equation}

\subsection{Simulation of point processes}

In this section, we present an overview of existing methods to generate pseudorandom numbers from a Poisson process, Non-homogeneous Poisson process and Hawkes process. Given predefined parameters of such processes, simulation is necessary to generate new data points, or events, from a point process. One can use these simulations in another model that requires random variables of stochastic nature. For instance, simulation of non-homogeneous process can be used as synthetic data for customers coming to a shop, or passengers coming to a bus stop \citep{kieu_stochastic_2018}. 

Simulation of homogeneous Poisson process is efficient and straightforward. First, we need a proposition that the number generated data points from a homogeneous Poisson process is also Poisson distributed. 

\begin{prop}
	For any interval of length $t$, the mean number of events generated by a homogeneous Poisson process is Poisson distributed with mean $\lambda t$ for all $s, t \geq 0$, 
	
	\begin{equation}
	P \{ N(t+s) - N(s) = n \} = e^{- \lambda t \frac{(\lambda t)^n}{n!}}
	\end{equation}
	
	for n=0,1,...
	
\end{prop}

This would directly lead us to the following proposition about the interarrival times

\begin{prop}
	The interarrival times $\{ Q_k \}_{k=1,2,..}$ of a homogeneous Poisson process with rate $\lambda>0$ are independent identically distributed exponential random variables with mean $ 1 / \lambda$. 
\end{prop} 

\textbf{Proof.} We have

\begin{equation}
P \{ Q_1 > t \} = P \{ N(t) = 0  \} = e^{-\lambda t} 
\end{equation}

so $Q_1$ is exponentially distributed with parameter $\lambda$. We use independent increments to estimate $Q_2$. 

\begin{equation}
P \{ Q_2 > t | Q_1 = a \} = P \{ N(a+b) -N(a) = 0 \} = e^{-\lambda t} 
\end{equation}

\begin{equation}
P \{ Q_2 > t \} = E [ P \{ W_2>t | W1  \}  ] = e^{-\lambda t} 
\end{equation}

so $W_2$ is exponentially distributed with parameter $\lambda$. The above equations also show that $W_2$ and $W_1$ are also independent. Repeating the same argument to the last interarrival times proves the proposition. 

The Algorithm \ref{algo:sim_HPP} can be used to generate the interarrival times $Q_k$ and simulate a homogeneous Poisson process \citep{ross_stochastic_1996}. 

\begin{algorithm} [H]	
	\SetAlgoLined
	\textbf{Input}: $\lambda(t), T$
	
	\textbf{Initialise} $n=0, t_0 = 0$;
	
	\While {\textit{True}}{
		Generate $ u \sim \text{uniform(0,1)}$;
		
		Let $Q = -\ln u / \lambda;$
		
		Set $t_{n+1} = t_n + Q$;
		
		\eIf{$t_{n+1}>T$}{
			return $\{ t_k\}_{k=1,2,...,n}$
		}{
		Set $n=n+1$
		}
	}
	
	\caption{Simulation of a Homogeneous Poisson process}
	\label{algo:sim_HPP}
\end{algorithm}

Simulating the points in an non-homogeneous Poisson process is more complicated. The points will need to be simulated sequentially due to the varying nature of the intensity function. In this chapter, we focus on describing the most popular method  for generating non-homogeneous Poisson process, which is the \textit{thinning} algorithm. The idea behind thinning is to first define a fixed rate function $\lambda_u(t) = \lambda_u$, which dominates the non-homogeneous function under study $\lambda(t)$. The next step is to generate from the implied homogeneous Poisson process with rate $\lambda_u$, and then reject an appropriate proportion of the simulated data points, such that the desired non-homogeneous Poisson process with rate $\lambda(t)$ is reached. However, simple acceptance-rejection thinning is prone to inefficiency. A straightforward enhancement to the thinning method is to reduce the excessive rejection by \textit{piecewise thinning}. Algorithm \ref{algo:sim_NHPP} describes this algorithm \citep{lewis_simulation_1979}.

\begin{algorithm} [H]	
	\SetAlgoLined
	\textbf{Input}: $\lambda(t), T$
	
	\textbf{Initialise} $n=m=0, t_0 = s_0 = 0, \hat{\lambda}=sup_{0 \leq t \leq T } \lambda(t)$;

	\While {$s_m < T$}{ 
		Generate $ u \sim \text{uniform(0,1)}$;
		
		Let $Q = -\ln u / \hat{\lambda};$
		
		Set $s_{m+1} = s_m + Q$;
		
		Generate $D \sim \text{uniform(0,1)}$;
		
		\If{$D \leq \lambda(s_{m+1}) / \hat{\lambda}  $}{
			$t_{n+1} = s_{m+1}$;
			$n=n+1$;
		}
	$m=m+1$;	
	}
	\eIf{$t_n \leq T $}{
		return $\{ t_k\}_{k=1,2,...,n}$
	
     }{
	
		return $\{ t_k\}_{k=1,2,...,n-1}$	
	}
	
	\caption{Simulation of a non-homogeneous Poisson process}
	\label{algo:sim_NHPP}
\end{algorithm}

To simulate the data points from a Hawkes process, let us look at the Equation \ref{dfn:Hawkes}. Given the first $k$ points ($t_1, t-2, ..., t_k$), the intensity $\lambda(t)$ is deterministic on [$t_k, t_{k+1}$]. Because the location of the next point $t_{k+1}$ is stochastic, generating the next point for a Hawkes process can be treated similar to generating the first point in a non-homogeneous Poisson process. Algorithm \ref{algo:sim_Hawkes} describes the modified thinning algorithm for generating data points from a Univariate Hawkes process with Exponential Kernel $\gamma(u)=\alpha e^{-\beta u}$ on [0,T], as proposed by \citet{ogata_lewis_1981}. 

\begin{algorithm} [H]	
	\SetAlgoLined
	\textbf{Input}: $\mu, \alpha, \beta, T$
	
	\textbf{Initialise} $\mathcal{T} = \varnothing, s=0, n=0$
	
	\While {$s < T$}{ 
		set $\hat{\lambda} = \lambda (s^+) = \mu + \sum_{\tau \in \mathcal{T} }^{max} \alpha e^{-\beta(s-\tau)}$;
		
		Generate $ u \sim \text{uniform(0,1)}$;
		
		Let $Q = -\ln u / \hat{\lambda};$
		
		Set $s = s + Q$;
		
		Generate $D \sim \text{uniform(0,1)}$;
		
		\If{$D \hat{\lambda} \leq \lambda(s) =  mu + \sum_{\tau \in \mathcal{T} }^{max} \alpha e^{-\beta(s-\tau)}$}{
			$n = n+1$;
			
			$t_n = s$;
			
			$\mathcal{T} = \mathcal{T} \cup \{ t_n\}$;
			
		}
		$m=m+1$;	
	}
	
	\eIf{$t_n \leq T $}{
		return $\{ t_k\}_{k=1,2,...,n}$
		
	}{
		
		return $\{ t_k\}_{k=1,2,...,n-1}$	
	}
	
	\caption{Simulation of a Hawkes process}
	\label{algo:sim_Hawkes}
\end{algorithm}

There are multiple ways to estimate the parameters of point processes, such as Maximum Likelihood Estimation or non-parametric methods such as Markov Chain Monte Carlo (MCMC) or Bayesian Inference. Interested readers may refer to \citet{diggle_statistical_2013} for more details regarding the inference of parameters of point processes from data.

\section{Methods and Tools to Analyse Point Patterns}

In this section we provide a discussion of methods and tools to analyse point patterns.

\subsection{Tools to Analyse Point Patterns}

There are a wide range of tools available for the analysis of point patterns, including bespoke packages written for statistical software, to modules written for Geographical Information Systems (GIS). Here we outline a selection of these and provide links to further reading. 

A number of packages exist for the R statistical language (see \cite{bivand_applied_2008} for a comprehensive discussion). The R package {\bf splancs} focuses on spatial point pattern analysis. The package {\bf spatstat} provides tools for analysing spatial point patterns in 2D and 3D, with some support for space-time analysis.  \citet{gabriel_stpp_2013} have produced the widely used R package \textbf{stpp} specifically for the analysis, simulation and display of spatial-temporal point processes. In the Python programming language, {\bf pointpats} (a sub-package of PySAL) deals with spatial point patterns.

The ArcGIS software programme provides a {\bf Spatial Statistics} toolbox which can be used to implement a range of analysis methods including GI* (see below) and for mapping spatial clusters. An open source GIS option is GeoDA.

%
%
\subsection{Visualising Point Patterns}

Visualising point patterns is an important means of attempting to understand their spatial pattern. However, simply displaying point data directly, particularly for data with large numbers of points, is not generally informative of the overall pattern. Points tend to overlap and it can be difficult for the brain to identify any patterns. Instead, it is common to use one of two approaches. The first is to produce a \textit{thematic maps}. To do so, the points ae overlaid with some area boundaries, the number of points per area are counted, and the areas are shaded according to the number of points that they contain. The second approach is to estimate the \textit{density} of the points. The `kernel density estimation' (KDE) algorithm is commonly used for this purpose. 

The KDE algorithm begins by placing a grid of  regularly-sized cells over the points. It then estimates the density of each cell in the grid by counting the number of cells within a distance (or `kernel'), $d$, away from the cell. \Citet{osullivan_geographic_2010} describe a ``naive'' method to estimate the density for a cell, $c$, as:
\begin{equation}
density_{c} = \frac { \#(S \in C(c,d)) } { \pi d^{2} }
\label{eqn:kde}
\end{equation}
where $d$ is the kernel distance, $C(c,d)$ is a circle of radius $d$ focused on the centre of the cell $c$, $S$ is the set of all points and $\#$ means the ``number of''. The size of the kernel ($d$) can be varied to find a balance between over-smoothing (where areas that have different densities are inadvertently merged into one large area)  and under-smoothing (where the output looks similar to the raw point pattern it began with). 

\begin{figure}[htbp] \begin{center} 
		\resizebox{1.0\textwidth}{!}{ 
			\includegraphics{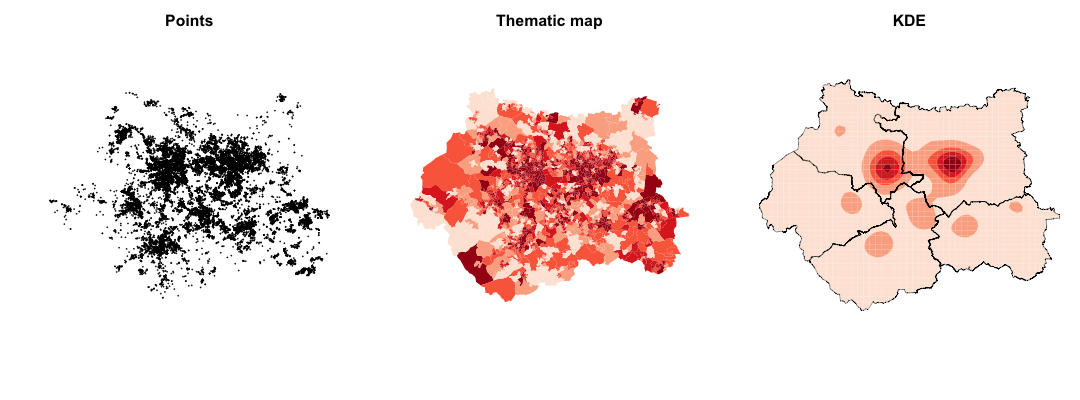}
		} 
		\caption{Visual exploration of point patterns: (a) displaying raw points; (b) a thematic map; (c) a density surface.} \label{fig:visual_comparison} \end{center} \end{figure}

To illustrate the output of these three methods,  Figure~\ref{fig:visual_comparison} compares a visualisation of points directly, a thematic map, and a density surface. Of these three methods, mapping the density is the most commonly used and often considered the most appropriate.

%
%
\subsection{Quantifying the Similarity between Point Patterns}

It is often useful to be able to quantify the spatial difference between two point patterns. This is, however, difficult with spatial data. Although not perfect, one way to quantify the difference between two patterns is to aggregate them to a spatial area -- i.e. overlay a spatial boundary dataset, such as a regular grid, and then count the number of points per area -- and thus create a table or matrix that captures the number of points from each dataset in each cell. Once this has been accomplished, goodness-of-fit statistics can be applied to quantify the difference between the two datasets. 

One of the easiest statistics to use is the residual sum of squares ($RSS$). This is calculated by adding the square of the differences between the counts in each cell:

\begin{equation}
RSS = \sum^{n}_{i=0}(y_{i}^{'} - y_{i} )^{2}
\end{equation}

where $y_{i}^{'}$ is the value from one of the datasets at in area $i$, and $y_{i}$ is the equivalent value from the other dataset. There are a number of disadvantages of the $RSS$, 
and hence there are a wealth of more advanced statistics whose aims are to give a more reliable assessment of the goodness of fit.

A drawback, however, with this approach is that the statistics are all a-spatial. Although they can be applied to spatial data through spatial aggregation, this aggregation introduces error. Also, the a-spatial statistics face difficulties where a cluster of points in two datasets are \textit{close} to each other, but not in exactly the same position. Fortunately, the combination of goodness-of-fit statistics along with other methods, such as those that interpret clustering, can provide additional rigour.

%
%

\subsection{Spatial Clustering}

The amount of clustering in a point pattern is an important property. Examining how the amount of clustering changes can provide information about the underlying process. For example, Figure~\ref{fig:point_pattern_randomness} illustrated how the degree of clustering can suggest whether a point pattern might be too clustered, or too uniform, to have originated by chance. A common means of quantifying the amount of clustering is to use nearest-neighbour methods. These methods inspect each point and compare the distance between it and its nearest neighbour (or, in some cases, a few neighbours). These distances provide evidence for how uniform or clustered a point pattern is. If distances are very small then the point pattern is probably clustered, if the distances are large then it is more likely to be uniform.

The mean minimum nearest neighbour distance, $\bar{d}_{min}$, in a point pattern is calculated by averaging the distances between each point, $i$, and its nearest neighbour, $j$:

\begin{equation}
\bar{d}_{min} = \sum_{i=1}^{n}\frac{d_{ij}}{n}
\end{equation}

where $n$ is the number of points. 

This \textit{mean minimum distance} -- i.e. the average of all minimum distances -- can then be used to calculate the Nearest Neighbour Index (NNI). The NNI is the ratio of the mean minimum distance observed from the point pattern to that which would have been generated if the points were produced by a random spatial process. It ranges from 0, indicating that all points are at the same location, to 2.15, suggesting entire spatial uniformity. One problem with the NNI is that it is susceptible to \textit{edge effects}. This occurs because points that are close to the edge of the pattern are likely to have larger minimum distances simply because there are fewer points near them. 



A more problematic drawback with index is that it only takes a single minimum distance into account and results in a single number that summarises the whole pattern. This disregards a lot of information about the pattern. An alternative nearest-neighbour statistic called Ripley's $K$~\citep{ripley_modelling_1977}, resolves this drawback. The $K$ function takes \textit{all} of the neighbours that are within a given distance from a point into account. The static can therefore be calculated for a number of different distances, $d$, with the value of $K$ at a distance $d$ ()$K(d)$) calculated as the mean of all counts divided by the overall point density. Larger values of $K(d)$ suggest that there is a greater amount of clustering at that distance than at others.

\begin{figure}[htbp] \begin{center} 
		\resizebox{1.0\textwidth}{!}{ 
			\includegraphics{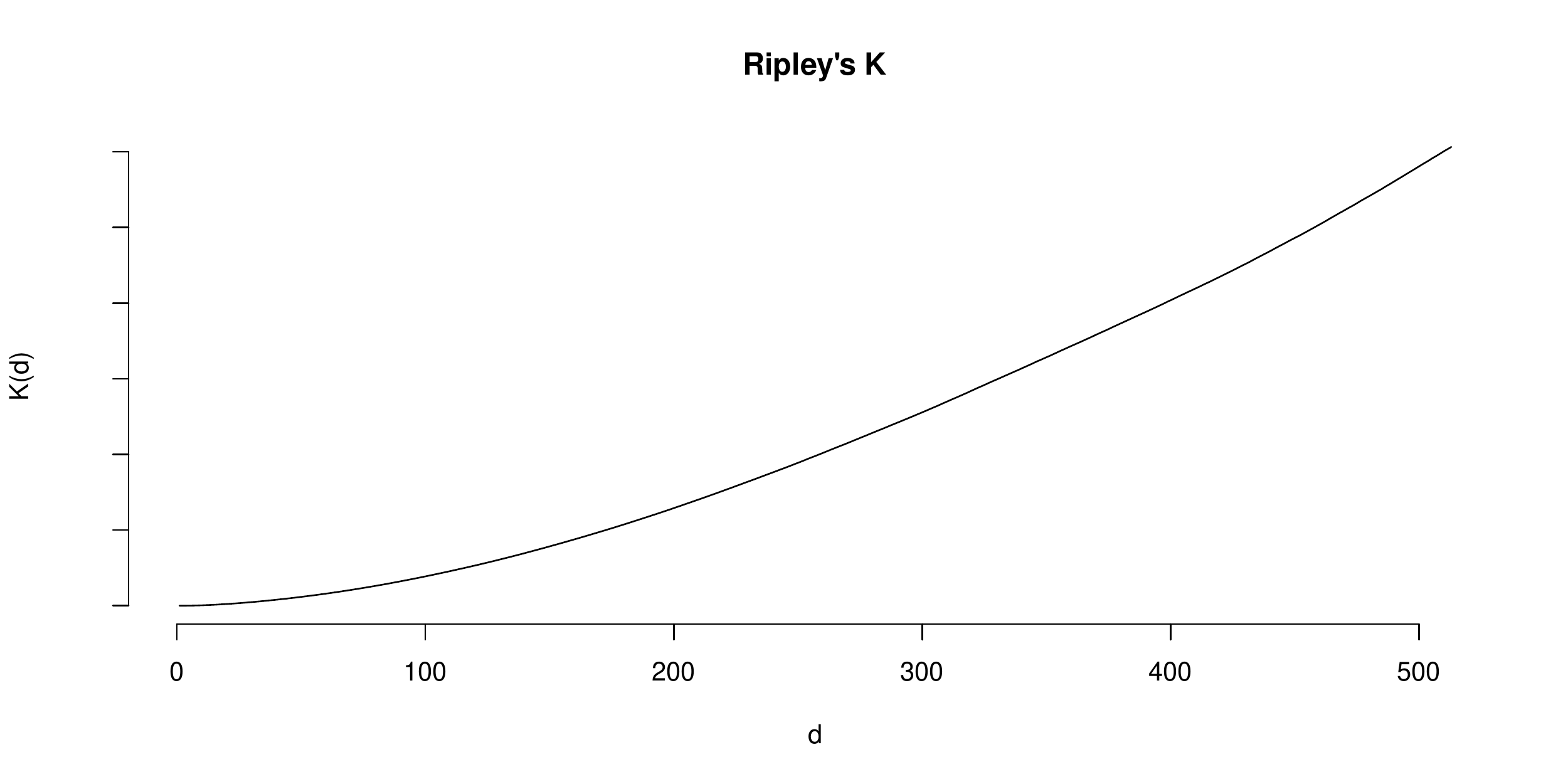}
		} 
		\caption{Ripley's $K$ with distance ($K(d)$) for two point patterns.} \label{fig:k_function}
	\end{center}
\end{figure}

The most useful output from the $K$ function come in the form of graphs with distance ($d$) on the horizontal axis and $K(d)$ on the vetrical. For example, Figure~\ref{fig:k_function} illustrates how $K(d)$ varies for an example point pattern. 

%
%
\subsection{Identifying the Location of Spatio-Temporal Clusters}

Clusters are defined as areas where the number of points in a local area is significantly greater than global averages~\citep{chainey_gis_2005}. The methods outlined in the previous section can be used to describe the overall amount of clustering in a point dataset, but provide no information about \textit{where} the clusters might be located. Two datasets might exhibit the same \textit{amount} of clustering, globally, but the actual clusters themselves might be located in different places. This section introduces two methods that can locate the clusters themselves; both spatially (using GI*) and temporally (using Spation-Temporal Scan Statistics). 

\subsection{Spatial Clusters: GI*} 

\begin{figure}[htbp] \begin{center} 
		\resizebox{1.0\textwidth}{!}{ 
			\includegraphics{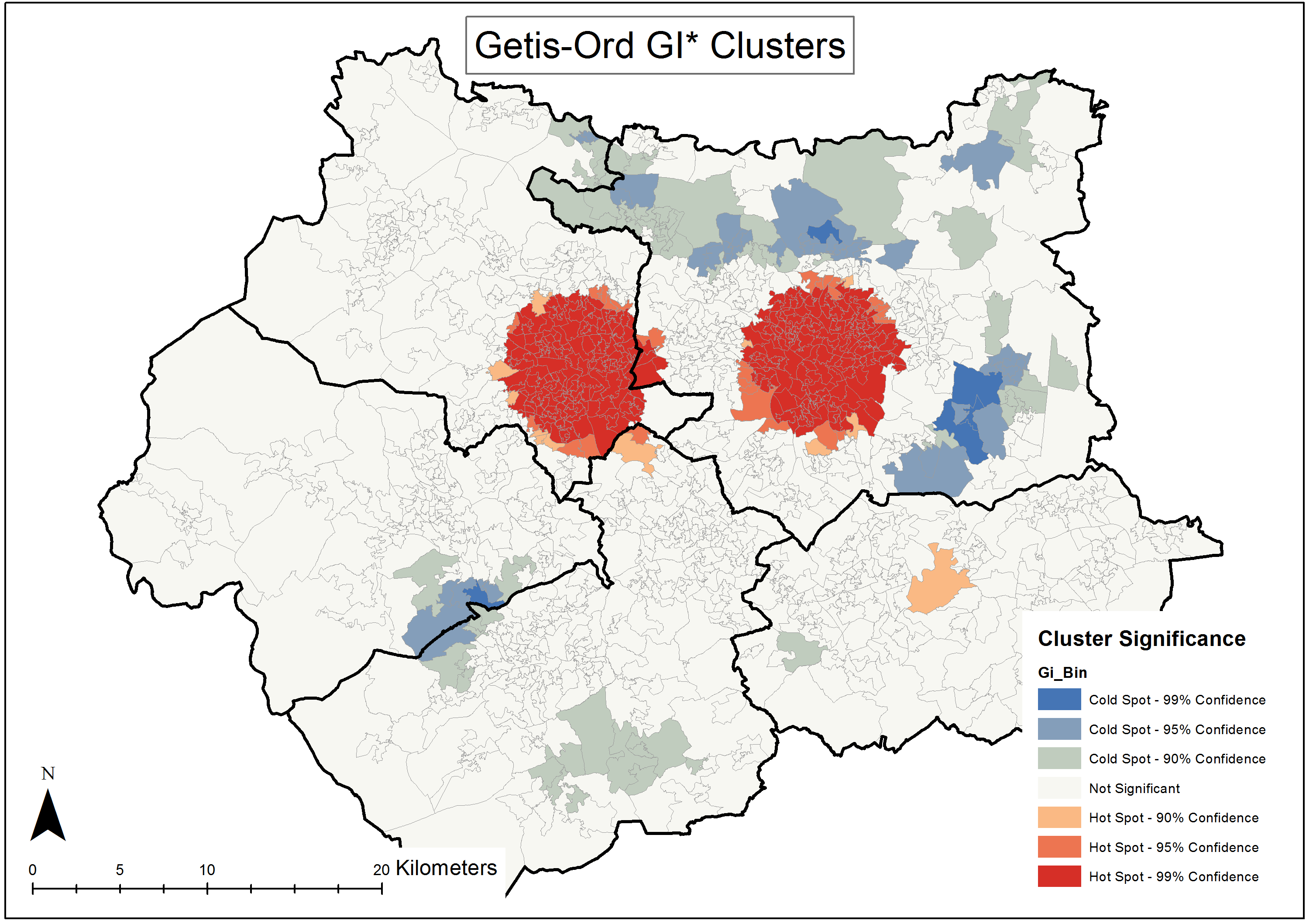}
		} 
		\caption{Getis-Ord GI* \citep{getis_analysis_1992, ord_local_1995} hotspots, produced using Arc Map software.} \label{fig:GIStar} \end{center} \end{figure}

The Getis-Ord GI* statistic \citep{getis_analysis_1992, ord_local_1995} estimates the spatial locations of areas that have a statistically significantly different number of points in them in comparison to their surroundings. It is an area-based measure, so point patterns first need to be aggregated to some areal boundaries -- administrative boundaries are commonly used, but abstract areas such as regular grids would be sufficient as well. The method provides a Z-Score in order to determine significance. For example, Figure~\ref{fig:GIStar} illustrates the GI* Z-Scores for the hypothetical point patterns used here. In this instance, the algorithm has identified the two most highly populated cities as significant hotspots. This is expected however, given the much larger population density in cities compared to their rural surroundings, so care must be taken to consider the size of the population at risk as well as the numbers of events.

\subsection{Spatio-Temporal Scan Statistics} 

An alternative to statistics such as GI* are \textit{spatial scan statistics}. These attempt to identify spatial clusters by moving a circular window of varying radius across a dataset and recording those circles that exhibit statistically significant numbers of events. The Geographical Analysis Machine \citep{openshaw_searching_1988} is an example of one such method. Although these are appropriate for identifying the locations of spatial clusters, they do not take the temporal dimension into account. In many applications -- such as epidemiology, disease control, crime analysis, etc. --  it might be necessary to identify the space \textit{and} time in which clusters occur. 

\begin{figure}[ht] \begin{center} 
		\resizebox{1.0\textwidth}{!}{ 
			\includegraphics{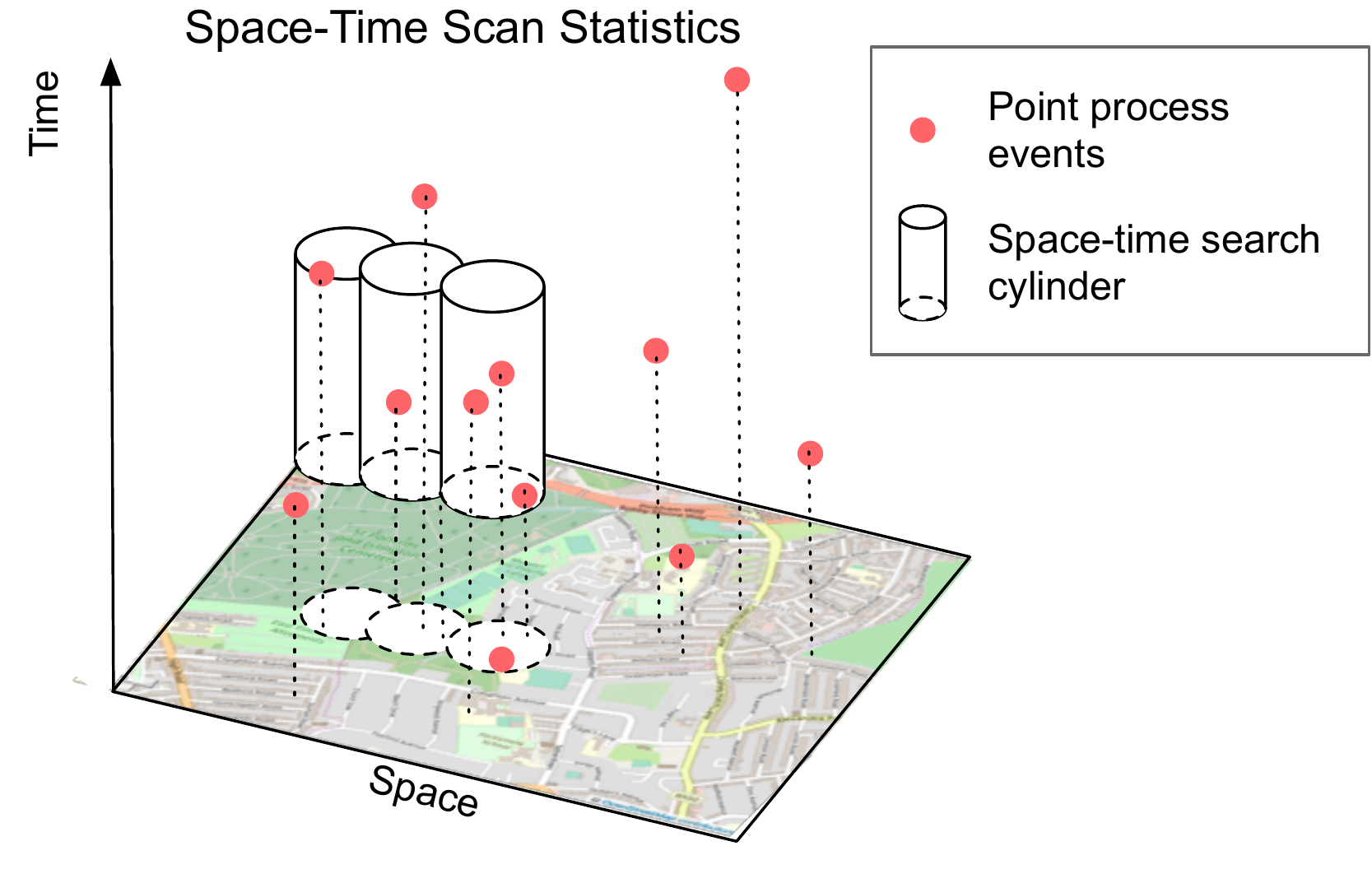}
		} 
		\caption{An example of searching for spatio-temporal clusters using a space-time cylinder. OpenStreetMap data is licenced under the Open Data Commons Open Database Licence (ODbL).} \label{fig:satscan}
	\end{center}
\end{figure}

Fortunately, spatial scan statistics can be adapted to take time into account by using a scanning window that covers a temporal dimension as well as the spatial dimension. Consider Figure~\ref{fig:satscan}. A Spatio-Temporal Scan algorithm can be perceived as a cylinder of varying radius and length. Like the circular window in a spatial scan, the spatio-temporal scan moves many cylinders through three dimensional data (with space on the horizontal axis and time on the vertical axis) recording the number of points that fall within the different cylinders. It is also possible to take a population-at-risk into account to identify clusters that are significant \textit{relative to a background population}. For example, an algorithm might assume that the number of events in a given space-time search cylinder follows a Poisson distribution. The null hypothesis s therefore that the expected number of events in each cylinder each point will be proportional to the population at risk at the same time and space.

\section{Conclusion}

This chapter has provided an overview of spatial, temporal and spatial-temporal point patterns and processes. We have discussed a range of models for analysis, including Homogeneous and non-Homogeneous Poisson processes and the Hawkes process. We also outline a range of methods and tools for analysing point patterns, including their visualisation, assessment of clustering, and extension to space-time statistics. There is a well established literature on the analysis of spatial or of temporal processes while spatial-temporal analysis is a relatively young field. Key texts we draw on in this chapter include \cite{diggle_statistical_2013},  \cite{diggle_spatiotemporal_2006} and \cite{daley_introduction_2003}. An increase in the availability of datasets which contain both the spatial and temporal location of events is driving the development of both methods and tools for analysis across a broad range of fields.

\bibliographystyle{chicago}
\bibliography{point_patterns}

\begin{thebibliography}{}

\bibitem[\protect\citeauthoryear{Bivand, Pebesma, and {G{\'o}mez-Rubio}}{Bivand
  et~al.}{2008}]{bivand_applied_2008}
Bivand, R., E.~J. Pebesma, and V.~{G{\'o}mez-Rubio} (2008).
\newblock {\em Applied Spatial Data Analysis with {{R}}}.
\newblock Use {{R}}! {New York, NY}: {Springer}.
\newblock OCLC: 254886734.

\bibitem[\protect\citeauthoryear{Chainey and Ratcliffe}{Chainey and
  Ratcliffe}{2005}]{chainey_gis_2005}
Chainey, S. and J.~Ratcliffe (2005).
\newblock {\em {{GIS}} and {{Crime Mapping}}}.
\newblock {Chichester, UK}: {John Wiley and Sons}.

\bibitem[\protect\citeauthoryear{Daley and {Vere-Jones}}{Daley and
  {Vere-Jones}}{2003}]{daley_introduction_2003}
Daley, D.~J. and D.~{Vere-Jones} (2003).
\newblock {\em An {{Introduction}} to the Theory of Point Processes}.
\newblock {New York}: {Springer}.
\newblock OCLC: 803632768.

\bibitem[\protect\citeauthoryear{Diggle}{Diggle}{2006}]{diggle_spatiotemporal_2006}
Diggle, P.~J. (2006).
\newblock Spatio-temporal point processes: Methods and applications.
\newblock {\em Monographs on Statistics and Applied Probability\/}~{\em 107},
  1.

\bibitem[\protect\citeauthoryear{Diggle}{Diggle}{2013}]{diggle_statistical_2013}
Diggle, P.~J. (2013, August).
\newblock {\em Statistical {{Analysis}} of {{Spatial}} and
  {{Spatio}}-{{Temporal Point Patterns}}\/} (3 edition ed.).
\newblock {Boca Raton}: {Chapman and Hall/CRC}.

\bibitem[\protect\citeauthoryear{Gabriel, Rowlingson, and Diggle}{Gabriel
  et~al.}{2013}]{gabriel_stpp_2013}
Gabriel, E., B.~Rowlingson, and P.~Diggle (2013).
\newblock {\textbf{Stpp}} : {{An}} {{{\emph{R}}}} {{Package}} for {{Plotting}},
  {{Simulating}} and {{Analyzing Spatio}}-{{Temporal Point Patterns}}.
\newblock {\em Journal of Statistical Software\/}~{\em 53\/}(2).

\bibitem[\protect\citeauthoryear{Getis and Ord}{Getis and
  Ord}{1992}]{getis_analysis_1992}
Getis, A. and J.~K. Ord (1992).
\newblock The {{Analysis}} of {{Spatial Association}} by {{Use}} of {{Distance
  Statistics}}.
\newblock {\em Geographical Analysis\/}~{\em 24\/}(3), 189--206.

\bibitem[\protect\citeauthoryear{Gonz{\'a}lez, {Rodr{\'i}guez-Cort{\'e}s},
  Cronie, and Mateu}{Gonz{\'a}lez et~al.}{2016}]{gonzalez_spatiotemporal_2016}
Gonz{\'a}lez, J.~A., F.~J. {Rodr{\'i}guez-Cort{\'e}s}, O.~Cronie, and J.~Mateu
  (2016, November).
\newblock Spatio-temporal point process statistics: {{A}} review.
\newblock {\em Spatial Statistics\/}~{\em 18}, 505--544.

\bibitem[\protect\citeauthoryear{Hawkes}{Hawkes}{1971}]{hawkes_spectra_1971}
Hawkes, A.~G. (1971).
\newblock Spectra of {{Some Self}}-{{Exciting}} and {{Mutually Exciting Point
  Processes}}.
\newblock {\em Biometrika\/}~{\em 58\/}(1), 83--90.

\bibitem[\protect\citeauthoryear{Kieu and Cai}{Kieu and
  Cai}{2018}]{kieu_stochastic_2018}
Kieu, L.~M. and C.~Cai (2018).
\newblock Stochastic collective model of public transport passenger arrival
  process.
\newblock {\em IET Intelligent Transport Systems\/}~{\em 12\/}(9), 1027--1035.

\bibitem[\protect\citeauthoryear{Lewis and Shedler}{Lewis and
  Shedler}{1979}]{lewis_simulation_1979}
Lewis, P. a.~W. and G.~S. Shedler (1979).
\newblock Simulation of nonhomogeneous poisson processes by thinning.
\newblock {\em Naval Research Logistics Quarterly\/}~{\em 26\/}(3), 403--413.

\bibitem[\protect\citeauthoryear{Openshaw, Charlton, and Craft}{Openshaw
  et~al.}{1988}]{openshaw_searching_1988}
Openshaw, S., M.~Charlton, and A.~Craft (1988).
\newblock Searching for {{Leukaemia Clusters Using}} a {{Geographical Analysis
  Machine}}.
\newblock {\em Papers in Regional Science\/}~{\em 64\/}(1), 95--106.

\bibitem[\protect\citeauthoryear{Ord and Getis}{Ord and
  Getis}{1995}]{ord_local_1995}
Ord, J.~K. and A.~Getis (1995).
\newblock Local {{Spatial Autocorrelation Statistics}}: {{Distributional
  Issues}} and {{An Application}}.
\newblock {\em Geographical Analysis\/}~{\em 27\/}(4), 286--306.

\bibitem[\protect\citeauthoryear{O'Sullivan and Unwin}{O'Sullivan and
  Unwin}{2010}]{osullivan_geographic_2010}
O'Sullivan, D. and D.~Unwin (2010, April).
\newblock {\em Geographic {{Information Analysis}}\/} (2nd Edition edition
  ed.).
\newblock {Hoboken, N.J}: {John Wiley \& Sons}.

\bibitem[\protect\citeauthoryear{Ozaki}{Ozaki}{1979}]{ozaki_maximum_1979}
Ozaki, T. (1979, December).
\newblock Maximum likelihood estimation of {{Hawkes}}' self-exciting point
  processes.
\newblock {\em Annals of the Institute of Statistical Mathematics\/}~{\em
  31\/}(1), 145--155.

\bibitem[\protect\citeauthoryear{Ripley}{Ripley}{1977}]{ripley_modelling_1977}
Ripley, B.~D. (1977).
\newblock Modelling {{Spatial Patterns}}.
\newblock {\em Journal of the Royal Statistical Society B\/}~{\em 39\/}(2),
  172--212.

\bibitem[\protect\citeauthoryear{Ross}{Ross}{1996}]{ross_stochastic_1996}
Ross, S.~M. (1996).
\newblock {\em Stochastic Processes\/} (2nd ed ed.).
\newblock Wiley Series in Probability and Statistics. {New York}: {Wiley}.

\bibitem[\protect\citeauthoryear{Schoenberg}{Schoenberg}{2003}]{schoenberg_multidimensional_2003}
Schoenberg, F.~P. (2003, December).
\newblock Multidimensional {{Residual Analysis}} of {{Point Process Models}}
  for {{Earthquake Occurrences}}.
\newblock {\em Journal of the American Statistical Association\/}~{\em
  98\/}(464), 789--795.

\end{thebibliography}

\end{document}